\documentclass[aps,prd,twocolumn,preprintnumbers,nofootinbib,showpacs,floatfix]{revtex4}

\usepackage{graphicx}
\usepackage{amsmath}

\newcommand{\MET}{{\slash\!\!\!\!E_T}}

\newcommand{\fbi}{\mathrm{fb}^{-1}}

\begin{document}

\title{Model independent reach for $W^\prime$ bosons at the LHC}

\author{Daniel~Duffty}
\email{dduffty@IIT.edu}
\affiliation{Department of Physics, Illinois Institute of Technology, Chicago, Illinois 60616-3793, USA}

\author{Zack~Sullivan}
\email{Zack.Sullivan@IIT.edu}
\affiliation{Department of Physics, Illinois Institute of Technology, Chicago, Illinois 60616-3793, USA}

\preprint{IIT-CAPP-12-07}

\begin{abstract}

  The semileptonic decay of single-top-quark production provides a
  strong probe for $W^\prime$ bosons at the CERN Large Hadron
  Collider.  We propose an explicit search strategy for $pp \to
  W^\prime \to tb \to l\nu bj$ for use at 7~TeV and 8~TeV collider
  energies, and integrated luminosities ranging from 5 to 20 $\fbi$.
  Based on detector-simulated results, we predict that a lower bound
  can be placed on the mass of right-handed $W^\prime_R$ with standard
  model-like couplings of $m_{W^\prime_R}>$1800 GeV at
  $\sqrt{S}=$7~TeV with 5~$\fbi$, and of $m_{W^\prime_R}>$2000~GeV at
  8~TeV with 20~$\fbi$.  For left-handed $W^\prime_L$ bosons we find a
  lower bound of 1750--1800 GeV at 7~TeV and 5~$\fbi$ depending on the
  sign of the interference with standard model single-top-quark
  production.  We present effective coupling $g^\prime$ dependent
  limits for accessible masses, and stress the importance of these
  limits for comparison with theoretical models.
\end{abstract}

\date{August 21, 2012}


\maketitle

%%%%%%%%%%%%%%%%%%%%%%%%%%%%%%%%%%%%%%%%%%%%%%%%%%%%%%%%%%%%%
\section{Introduction}
\label{sec:intro}

The search for new charged vector currents, generally called
$W^\prime$ bosons, plays an important role in many extensions of the
standard model.  Some theories propose higher mass $W$ boson
resonances \cite{Datta:2000gm}; while others propose right-handed
counterparts to the left-handed standard model $W$ in a broken
$SU(2)_L \times SU(2)_R$ symmetry
\cite{Pati:1974yy,Mohapatra:1974hk,Mohapatra:1974gc,Senjanovic:1975rk}. Still
others propose a heavy mass eigenstate in strongly interacting
theories, such as non-commuting extended technicolor
\cite{Chivukula:1995gu}.  All of the $W^\prime$ bosons in these
theories enter the Lagrangian with terms of the form
\begin{equation}
\label{eq:L}
\mathcal{L}=\frac{g'}{2\sqrt{2}} V^\prime_{ij} W^\prime_\mu \bar f^i \gamma^\mu 
(1\pm\gamma_5) f^j+\mathrm{H.c.}\,,
\end{equation}
which mirrors that of the standard model $W$ (without the lepton
sector if it is right-handed).  While there could be left-right
mixing, such mixing is constrained by $K$--$\overline{K}$ mixing
\cite{Groom:in}; hence, theories have been proposed that would suppress
this naturally via orbifold breaking of the left-right symmetry
\cite{Mimura:2002te} or supersymmetric interactions
\cite{Cvetic:1983su}.

While the phenomenology of models beyond the standard model (SM) is
generally complex, it was demonstrated in Ref.\ \cite{Sullivan:2002jt}
that $W^\prime$ sectors can be factorized through next-to-leading
order (NLO) in QCD into terms of the form of Eq.\ \ref{eq:L}.  Hence,
it was proposed that searches for direct resonances decaying into a
$tb$ final state be used to bound all such possible models that couple
to quarks.  Following Ref.\ \cite{Sullivan:2002jt}, a series of
searches were performed by the CDF
\cite{Acosta:2002nu,Aaltonen:2009qu}, and D0
\cite{Abazov:2006aj,Abazov:2011xs} Collaborations at the Fermilab
Tevatron, setting the world's strongest bounds on right-handed
$W^\prime$ bosons, and competitive bounds on left-handed $W^\prime$
bosons \cite{Beringer:1900zz}.

Recent interest by the ATLAS and CMS Collaborations in extending these
studies to the CERN Large Hadron Collider (LHC), has prompted a
re-examination of the reach and interpretation of these models.  Early
results by ATLAS \cite{Aad:2012ej} and CMS \cite{:2012sc} utilize NLO
cross sections from an early draft of this paper (reproduced here in
the Appendix).  In Sec.\ \ref{sec:results} we compare our predictions
with these first results.

In this paper we extend earlier predictions for the model-independent
reach for $W^\prime$ bosons at a 14~TeV LHC \cite{Sullivan:2003xy} to
7~TeV and 8~TeV energies.  In Sec.\ \ref{sec:sim} we provide details
of our detector simulation.  We point out previously undescribed
kinematic differences between right- and left-handed $W^\prime$ bosons
($W^\prime_{R,L}$) that play a role in the reach at the LHC.  In
addition, we examine the effect of $W^\prime$ charge on the
distributions at a $pp$ collider.  In Sec.\ \ref{sec:results} we
propose a set of cuts for the model-independent analysis, and describe
our predictions for the reach at the LHC for 7~TeV and 8~TeV.  We
conclude with suggestions for further research.  In the Appendix, we
provide updated predictions for NLO $W^\prime$ cross sections for
7~TeV and 8~TeV $pp$ colliders, including all theoretical
uncertainties, for use by the coming experimental analyses.

%%%%%%%%%%%%%%%%%%%%%%%%%%%%%%%%%%%%%%%%%%%%%%%%%%%%%%%%%%%%%%%%%%%%%%%%%%
\section{Simulation}
\label{sec:sim}

The signal of interest is $pp \to W^\prime \to t b$, where the top
quark decays as $t\to Wb$, and the $W$ decays leptonically to an
electron or muon plus a neutrino.  In order to simulate the full
decay chain, including all angular correlations, we utilize a general
$W^\prime$ model \cite{ZhouSullivan} in \textsc{MADEVENT}
\cite{Alwall:2007st} to produce parton-level signal and background
events.  These events are fed through \textsc{PYTHIA}
\cite{Sjostrand:2006za} to generate initial- and final-state
showering, and reconstructed in an ATLAS-like detector model in a
modified \textsc{PGS} 4 \cite{conway:2006pgs} detector simulation.

We add an anti-$K_T$ jet reconstruction algorithm to \textsc{PGS} in
order to match the current jet algorithms used by the ATLAS and CMS
Collaborations.  We use a cone size of $0.4$ for the anti-$K_T$
cutoff, and apply a jet energy scale (JES) correction to recover
energy lost due to detector resolution and limited cone size.  This
JES correction is extracted from identified $b$ jets in a $Zb\to
e^+e^- b$ test sample.  The resulting correction is small for jet
energies grater than $\sim 50$ GeV, and is implemented by scaling the
jet four-momentum as
\begin{equation}
\label{Eq:jec}
p_\mu^\prime = p_\mu\left( 1+ \frac{2.2}{E} + \frac{62.2}{E^2} \right) \,. 
\end{equation}
After the jet energy correction is applied, we find very little
dependence in the final results on cone sizes between $0.4$--$0.7$.

An important feature in reducing backgrounds to the $tb$ final state
is the use of $b$ tagging.  We model $b$-tagging with the default
\textsc{PGS} tight-tagging algorithm, modified to include muon tracks
inside of a jet.  For the cuts we employ below this leads to $\sim
50\%$ $b$-tagging efficiency, with a somewhat over-estimated $\sim
2\%$ mistag rate for light quarks, and a charm mistag rate of $\sim
10\%$.  The effect of other $b$-tagging scenarios is addressed in
Sec.\ \ref{sec:results}.

Since our interest is in $W^\prime$ bosons with masses near or above
1~TeV, we expect the $b$-jet that recoils against the top quark in the
event to have a transverse energy $E_T\sim 500$--$1200$ GeV.  At this
stage it is unclear what the ultimate $b$-tagging efficiency will be
for these high energy jets, however, we expect the decay products of
the $B$ hadrons to be so boosted that secondary vertex reconstruction
will be difficult.  For the purposes of this study we assume that we
are unable to make use of $b$-tagging for the leading jet in our
events.  Should an algorithm for high efficiency and high purity $b$
jets near 1~TeV be developed, it could improve the signal purity.

The primary backgrounds of this $W^\prime$-induced $s$-channel
single-top-quark process will be $t$-channel single-top-quark
production, $t\overline{t}$ decaying to a lepton plus jets, and $Wjj$
production.  $t$-channel single-top-quark production is significant
because the lead jet and the top quark will have a large invariant
mass due to the large angular separation between decay products.
$Wjj$ is a major background strictly because its large cross section
compensates for its small light jet mistag rate.  Other important
backgrounds are the dilepton decay channel of $t\overline{t}$, where
one of the leptons is lost within a jet, other $W+$jet events ($Wcj,
Wcc, Wbj, Wbb$), and standard model $s$-channel single-top-quark
production.

Background events are reweighted to match their NLO cross sections
using a scale of 1~TeV calculated using \textsc{MCFM} \cite{Campbell:2010ff}
after acceptance cuts, and CTEQ 6.6 parton distribution functions
\cite{Nadolsky:2008zw}.  The $s$- and $t$-channel single-top-quark
normalizations are confirmed with matching to \textsc{ZTOP}
\cite{ztop}.  Normalizing the $W^\prime$ signal to NLO is somewhat
more subtle.  Here we use the code from Ref.\ \cite{Sullivan:2002jt}
updated for 7 and 8~TeV $pp$ colliders (see the Appendix for
inclusive cross sections with theoretical errors).

The normalization for right-handed $W^\prime_R$ bosons is
straight-forward, but left-handed $W^\prime_L$ bosons can interfere
with standard model $W$-propagated single-top-quark production.  As
pointed out in Ref.\ \cite{ZhouSullivan}, the $W^\prime$--$W$
interference can be constructive, destructive, or negligible,
depending on the sign and size of the $V^\prime_{tb}$ term in the
$W^\prime$ mixing matrix with respect to the other elements; a
positive term (as is usually assumed \cite{Boos:2006xe}) in the mixing
matrix yields a destructive interference, while a negative term in the
mixing matrix provides a constructive interference.  We consider the
two limiting cases of fully destructive and constructive interference
below in order to bound the range of possible results.

\begin{figure}
\includegraphics[width=\columnwidth]{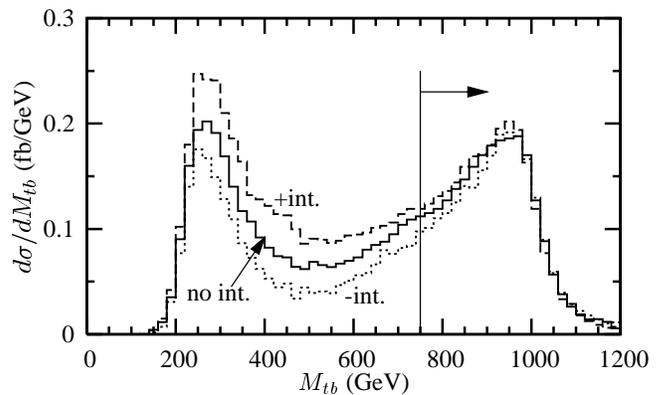}
\caption{Reconstructed invariant mass $M_{tb}$ for a 1~TeV SM-like left-handed
$W^\prime$ boson and standard model $pp\to W\to tb$ $s$-channel production
processes.  These processes can be non-interfering,
constructively interfering ($+$ int.), or destructively interfering ($-$ int.),
but at the large invariant mass relevant for discovery the interference
is small.
\label{fig:NonConDesCS}}
\end{figure}

To fix the NLO normalization for left-handed $W^\prime_L$ bosons, we
extract a $K$-factor after cuts from the case of no interference, and
scale up or down the events in the interfering cases by the same
$K$-factor.  We justify and quantify the error in this approximation
as follows: First, we observe in Fig.\ \ref{fig:NonConDesCS} that for
maximal SM-like coupling ($g^\prime/g_{SM} = 1$) the interference can
be large in certain regions of reconstructed invariant mass of the
$tb$.  Fortunately, we are only interested in large invariant masses
--- close to the $W^\prime$ boson mass --- where the interference is
never more than $\sim 20\%$.  The $K$-factor itself is typically $\sim
1.2$ for the masses we consider.  Furthermore, both standard model
single-top-quark production and $W^\prime_L$ production have
identically factorizable matrix amplitudes \cite{Sullivan:2002jt} at
NLO.  Hence, they receive the same QCD corrections at NLO.  Put
together, we estimate the maximum error we introduce by using
leading-order interference and normalizing in this fashion is $\sim
0.2\times 0.2$ or 4\%.  This error is negligible when compared with
the 10--30\% error introduced by parton distribution function
uncertainties (listed in the Appendix).

%%%%%%%%%%%%%%%%%%%%%%%%%%%%%%%%%%%%%%%%%%%%%%%%%%%%%%%%%%%%%%
\subsection{Kinematic features}
\label{sec:kinematics}

The inclusive cross sections for right- and left-handed $W^\prime$
bosons decaying to $tb$ differ mostly in their branching fractions.
Since there are no light right-handed neutrinos, we expect roughly
$4/3$ as many $tb$ from $W^\prime_R$ decay as from $W^\prime_L$ decay.
The larger branching fraction suggests there will be a slightly better
reach for right-handed bosons than left-handed bosons.  This is true,
but there are kinematic differences between right- and left-handed
bosons, as well as between $W^{\prime +}$ and $W^{\prime -}$ bosons,
that affect the acceptances and reconstruction efficiencies.

Spin correlations, usually considered as a way to distinguish
right-handed from left-handed $W^\prime$ bosons, also modify the
distributions of jets and leptons in the detector.  The dominant
parton luminosity for $W^\prime$ boson production involves a
valence-sea quark combination, $u\bar d$ or $d\bar u$.  On average
this leads to a forward boost of the $W^\prime$ bosons and their decay
products in the direction of the valence quark.  Spin correlations
between the down-type parton in the initial state and the $b$-jet from
the top quark decay, or the $d$ and the charged lepton from the $W$
decay affect right- and left-handed $W^\prime$ bosons differently.

\begin{figure}
\includegraphics[width=\columnwidth]{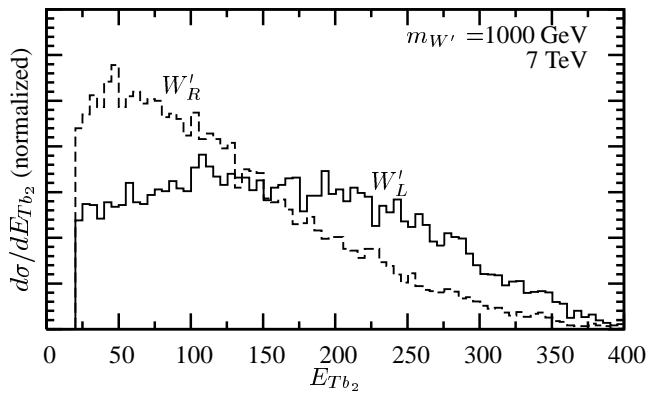}
\caption{Normalized transverse energy distribution of the second $b$
  jet $E_{Tb_2}$ for $W^\prime_R$ and $W^\prime_L$ bosons with a mass
  of 1~TeV.}
\label{fig:EtBLR}
\end{figure}

\begin{figure}
\includegraphics[width=\columnwidth]{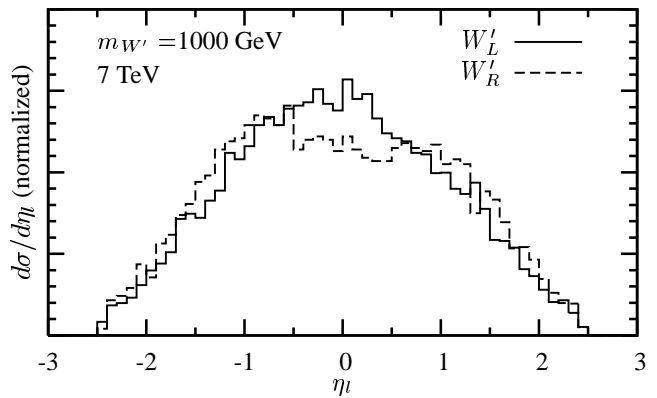}
\caption{Normalized lepton pseudorapidity $\eta_l$ for $W^\prime_R$ 
and $W^\prime_L$ bosons with a mass of 1~TeV.}
\label{fig:EtaLepLR}
\end{figure}

Bottom jets from the top quark in $W^\prime_R$ decay are partially
antialigned with the $W^\prime_R$ direction, leading to a slightly
softer transverse energy $E_{T b}$ spectrum than for $W^\prime_L$
bosons.  In Fig.\ \ref{fig:EtBLR} we see this softer $E_{Tb}$ spectrum
for 1~TeV $W^\prime_R$ vs.\ $W^\prime_L$ bosons.  This feature reduces
our final prediction of the reach for right-handed $W^\prime$ bosons
at lower masses, as a large fraction of events fail to pass minimal
jet acceptance cuts.  Lepton acceptance is also reduced for
$W^\prime_R$ bosons, as the spin correlations make them more forward;
though we see in Fig.\ \ref{fig:EtaLepLR} the effect is small.

\begin{figure}
\includegraphics[width=\columnwidth]{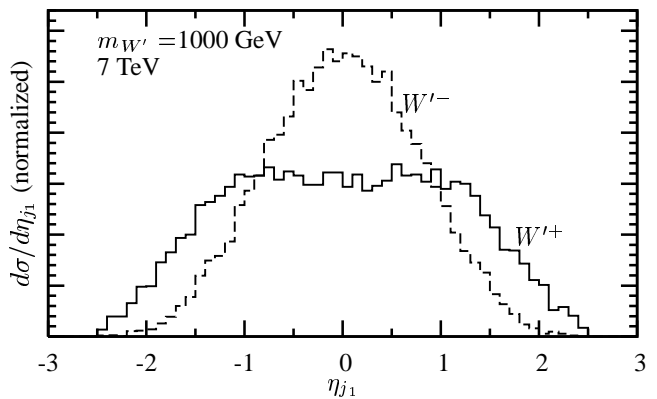}
\caption{Normalized pseudorapidity distribution of the leading jet in
$W_R^{\prime +}$ and $W_R^{\prime -}$ decays for a $W^\prime$ mass of 1~TeV.
\label{fig:eta_jet+-}}
\end{figure}

\begin{figure}
\includegraphics[width=\columnwidth]{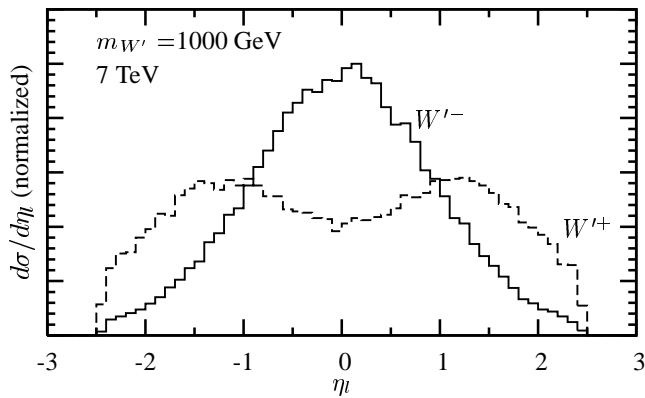}
\caption{Normalized pseudorapidity distribution of the charged lepton in
$W_R^{\prime +}$ and $W_R^{\prime -}$ decays for a $W^\prime$ mass of 1~TeV.
\label{fig:eta_lep+-}}
\end{figure}

More striking than left-right differences, are the differences between
$W^{\prime +}$ and $W^{\prime -}$ bosons.  The cross section for
$W^{\prime +}$ is roughly twice that of $W^{\prime -}$, since there
are roughly twice as many valence $u$ quarks as valence $d$ quarks in
the proton.  The spin correlations exaggerate the effect to produce
very different rapidity distributions for the final state particles.
The leading jet in the $W^{\prime -}$ decay is more central in
pseudorapidity $\eta_{j_1}$ than in $W^{\prime +}$ decay, as shown
in Fig.\ \ref{fig:eta_jet+-} for right-handed $W^\prime$ bosons.
Fortunately, detector acceptance at the LHC covers the entire rapidity
range for both.  The lepton pseudorapidity $\eta_l$ in Fig.\
\ref{fig:eta_lep+-}, however, is more forward for $W^{\prime +}$
than for $W^{\prime -}$.  This will lead to a slightly different
detector response between the two production modes.  In this analysis
we sum over both $W^{\prime +}$ and $W^{\prime -}$ production with the
same cuts, but future studies might consider optimizing cuts for
$W^{\prime +}$ and $W^{\prime -}$ analyses separately.

%%%%%%%%%%%%%%%%%%%%%%%%%%%%%%%%%%%%%%%%%%%%%%%%%%%%%%%%%%%%%%%%%%%%%%%%%%
\section{Results}
\label{sec:results}

We propose a simple cut-based search for $W^\prime$ bosons that is
effective at 7~TeV and 8~TeV energies at the LHC.  While we saw in
Sec.\ \ref{sec:sim} several strong angular correlations that affect
acceptances, we find that the ultimate significance of a $W^\prime$
signal in the presence of backgrounds is dominated by purely kinematic
effects.  Hence, we make use of those distinctive kinematic features
here.

In this analysis we are examining the $pp\to W^\prime \to tb \to l\nu
bj$ final state, where $l$ is an electron or muon.  Since all of our
signal events are fairly central, we begin with a basic set of
detector acceptance cuts that require at least two jets with $E_{T
  j}>20$ GeV and $|\eta_j|<2.5$, one lepton with $p_{T l}>20$ GeV and
$|\eta_l|<2.5$, and missing transverse energy $\MET>20$ GeV.  (All
cuts are summarized in Table \ref{tab:cuttable}.)  At the level of
acceptance cuts the signal to background ratio $S/B\sim 1/1000$ for a
canonical right-handed $W^\prime$ boson with standard model-like
couplings ($g^\prime/g_{SM} = 1$).

\begingroup
\begin{table}[!htb]
  \caption{Acceptance and analysis cuts for $pp\to W^\prime \to tb \to 
l\nu b j$.
\label{tab:cuttable}}
\begin{ruledtabular}
\begin{tabular}{lll}
Lead jet& $E_{Tj_1} > 0.2 m_{W^\prime}$ & $|\eta_{j_1}| < 2.5$\\
$b$-tagged jet& $E_{Tb} > 20$ GeV & $|\eta_{b}| < 2.5$ \\
Leading $e^\pm$ or $\mu^\pm$ & $p_{Tl_1} > 20$ GeV & $|\eta_{l_1}| < 2.5$\\
Second $e^\pm$ or $\mu^\pm$ & $p_{Tl_2} <  10$ GeV;\  or & $|\eta_{l_2}| > 2.5$\\
Missing $E_T$ &  $\slash\!\!\!\!E_T > 20$ GeV & \\
Reconstructed top &  $M_{l\nu b} < 200$ GeV & \\
$W^\prime$ mass window &  \multicolumn{2}{l}{$0.75 m_{W^\prime} < M_{l\nu bj} < 1.1 m_{W^\prime}$}
\end{tabular}
\end{ruledtabular}
\end{table}
\endgroup

The most distinctive feature of these $W^\prime$ bosons is the highly
energetic leading bottom jet.  As mentioned above, we avoid attempting
to $b$-tag this jet, and instead simply require it to have a
transverse energy $E_{T j_1}> 0.2 m_{W^\prime}$.  This cut has a
minimal effect on the signal, since the leading jet has energies
approaching $0.5m_{W^\prime}$ (up to detector resolution).  However,
all backgrounds have a leading jet $E_T$ that is falling with energy
for the masses we consider.  As we see in Table \ref{tab:cuteffect},
after this cut $S/B$ improves to $1/20$ for a 1~TeV SM-like
$W^\prime_R$.  The dominant background is due to mistags from $Wjj$,
but this is reducible.

\begingroup
\begin{table}[!htb]
\caption{Cross sections (fb) for signal and backgrounds at each level of cuts,
assuming a 1~TeV right-handed $W^\prime$ boson with standard model-like
couplings ($g^\prime/g_{SM} = 1$) at a 7~TeV LHC.
\label{tab:cuteffect}}
\begin{ruledtabular}
\begin{tabular}{rrrrrr}
Process & $jjl\,\MET$ & $E_{T j_1}\,\textrm{cut}$ & $b\,\textrm{tag}$ & $M_{l\nu b}$ & $M_{l\nu bj}$ \\\hline
$Wjj$, $Wcc$, $Wcj$ & 219000 & 5680 & 230 & 83.8 & 12.9\\
$Wbb$, $Wbj$ & 2580 & 42.3 & 16.4  & 6.4 & 0.8\\
$t\overline{t}$ & 8010 & 136 & 70.2  & 40.3 & 8.4\\
$t$-chan.\ single top & 1590 & 61.3  & 30.1 & 23.8 & 6.8\\
$s$-chan.\ single top & 182 & 8.5  & 3.6  & 2.7 & 0.4\\\hline
Background total & 231000 & 5830 & 350 & 157 & 29.3\\
$W^\prime$ signal& 294 & 247 & 106  & 79.8 & 76.3\\
\end{tabular}
\end{ruledtabular}
\end{table}
\endgroup

The next most distinctive element of the signal is the $b$ jet coming
from the top quark decay.  This jet is often, but not always, the
second-hardest jet in the event.  Sometimes showered jets accidentally
have a larger energy, and sometimes the jet that recoils against the
top quark is reconstructed at lower energy.  In order to capture most
of the signal events, we require at least one $b$-tagged jet that is
\textit{not} the leading jet in the event.  If there are more than one
$b$-tagged jet, we assume the highest $E_T$ $b$-jet is the one coming
from the top quark decay.  The main effect of this cut is to reduce
the $Wjj$-oriented backgrounds ($Wjj$, $Wcj$, and $Wcc$) by a factor
of 20, improving $S/B$ to roughly $1/3.5$ for a SM-like $W^\prime_R$.

The relative sizes of the $Wjj$, $t$-channel single top quark, and $t
\bar t$ backgrounds are highly affected by the choice of $b$-tagging
efficiencies.  A large $b$ acceptance rate would allow a greater
acceptance of signal events, but would have a relatively larger
proportion of $Wjj$ events.  For example, a 70\% $b$ acceptance is
achievable \cite{Aad:2009wy} and would increase the signal and
background top-quark final states acceptance by about 40\%, but the
$Wjj$ backgrounds would increase by more than a factor of 2 with
current algorithms.  The net effect would be a lower signal purity,
and no gain in significance.

For this analysis we choose to improve the signal purity by roughly
reconstructing a top-quark mass out of the $b$-tagged jet, the lepton,
and the missing energy.  We first reconstruct the $W$ in top decay by
assuming the $W$ is on-shell, and choosing the smallest rapidity
solution for the neutrino four-momentum.  In order to suppress
sensitivity to the jet energy resolution, we place a mild upper cut on
the $l\nu b$ invariant mass of $M_{l\nu b}<200$ GeV.  By choosing to
ignore $b$ tags of the leading jet, this cut reduces the $t\bar t$
background, as there is a 50\% chance of tagging the $b$ jet that is
not associated with the leptonic final state.  This cut is useful in
obtaining the strongest limit on the $W^{\prime} q\bar q$ coupling
$g^\prime$, but we explain in the Conclusions why this cut might be
removed for more general studies.

\begin{figure}
\includegraphics[width=\columnwidth]{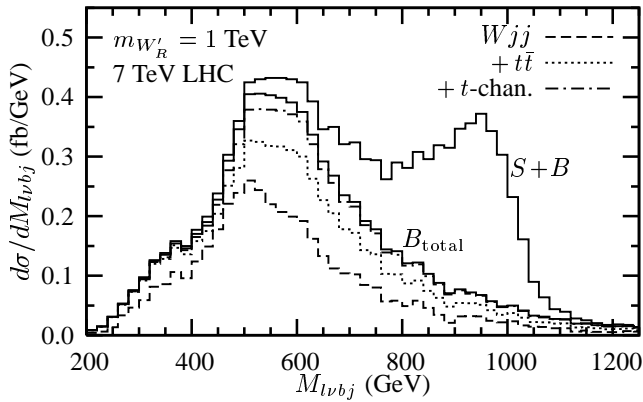}
\caption{Differential cross section in the reconstructed invariant
mass $M_{l\nu bj}$ for the signal $S$ and backgrounds, for a SM-like
right-handed $W^\prime_R$ boson of mass of 1~TeV after acceptance cuts.
The total background $B$ is composed mostly of $Wjj$, $t\bar t$, and
$t$-channel single-top-quark production.
\label{fig:1000backgrounds}}
\end{figure}

After the cuts considered so far, a 1~TeV right-handed $W^\prime$
boson with SM-like couplings would have a very strong signature at the
LHC.  In Fig.\ \ref{fig:1000backgrounds} we see the cross section as a
function of $M_{l\nu bj}$ invariant mass for the signal plus
background compared to the steeply falling background.  The background
under the $W^\prime$ mass peak is composed of nearly equal parts
$t$-channel single-top-quark production, $t\bar t$, and $Wjj$.  In
this analysis we consider $W^\prime$ masses from 500 GeV to 3.5~TeV,
and find that most of the signal events tend to fall in a mass window
of $0.75 m_{W^\prime} < M_{l\nu bj} < 1.1 m_{W^\prime}$.  When this
cut is applied to the 1~TeV SM-like $W^\prime_R$ of
Table~\ref{tab:cuteffect}, we see that $S/B$ improves from $1/2$ to
nearly $2.6/1$, and the significance for discovery ($S/\sqrt{S+B}$) is
greater than 16 with 5~$\fbi$ of integrated luminosity at 7~TeV.  This
particular $W^\prime$ has already been excluded by early LHC data
\cite{:2012sc,Aad:2012ej}, but we use these cuts to answer what is the
reach in effective coupling $g^\prime/g_{SM}$ vs.\ $W^\prime$ boson
mass.

\begin{figure}
\includegraphics[width=\columnwidth]{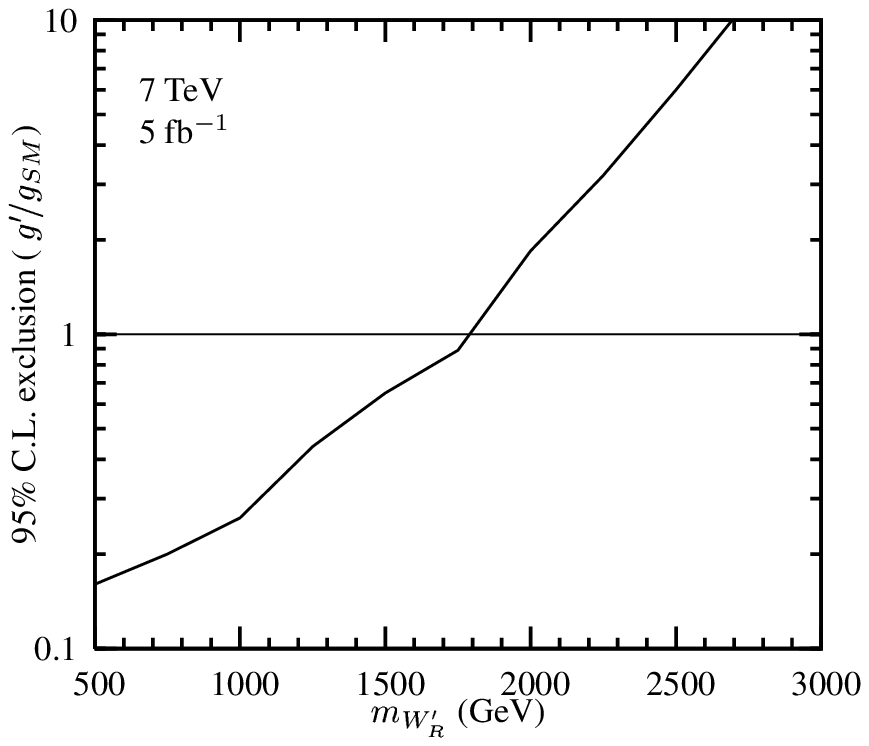}
\caption{Expected 95\% C.L.\ exclusion limit for $g^\prime/g_{SM}$ for a
  right-handed $W^\prime_R$ boson at $\sqrt{S}=7$ TeV and 5 fb$^{-1}$ of data.
\label{fig:RequiredCoupRS}}
\end{figure}

\begin{figure}
\includegraphics[width=\columnwidth]{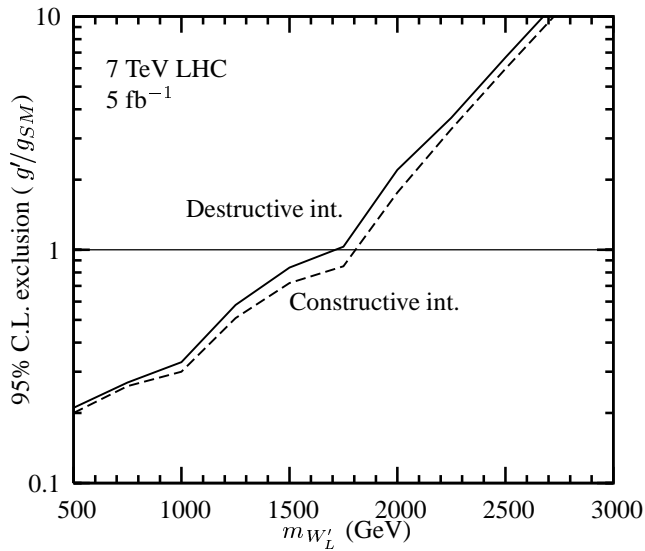}
\caption{Expected 95\% C.L.\ exclusion limit for $g^\prime/g_{SM}$ for a 
  left-handed $W^\prime_L$ boson at $\sqrt{S}=7$ TeV and 5 fb$^{-1}$ of data
  that interferes constructively or destructively with standard model
  single-top-quark production.
  \label{fig:RequiredCoupLS}}
\end{figure}

In Table~\ref{tab:7coupling} we list the 95\% confidence level (C.L.)
exclusion reach in effective coupling $g^\prime/g_{SM}$ with
5~fb$^{-1}$ of integrated luminosity for masses between 500--3000 GeV
at 7~TeV.  This ratio is most useful for direct comparison to
theoretical models with Lagrangians of the form in Eq.\ \ref{eq:L}
\cite{Sullivan:2002jt,Sullivan:2003xy,ZhouSullivan}.  For right-handed
$W^\prime_R$ searches the reach is approximately $\sqrt{4/3}$ times
better than for left-handed $W^\prime_L$ due to the larger branching
fraction for $W^\prime_R\to tb$.  As can be seen in Figs.\
\ref{fig:RequiredCoupRS} and \ref{fig:RequiredCoupLS}, $W^\prime$
bosons can be excluded for $g^\prime/g_{SM}$ down to a few times
$10^{-1}$ below 1.5~TeV.  This is significantly smaller than the
$g^\prime$ that appear the models described in Sec.\ \ref{sec:intro}.
For standard model-like $W^\prime$ bosons, a limit could be set around
$1.7$--$1.8$ TeV, depending on handedness and the sign of
interference.  Note that the difference between constructive and
destructive interference is only about a $\sim 10\%$ effect on the
final limit at any given mass.  It is also important for experiments
to demonstrate the exclusion for models with $g^\prime/g_{SM} > 1$, as
models remain perturbative up to a ratio of about 5.  For example,
Kaluza-Klein models can have ratios of $\sqrt{2}$ or 2
\cite{Mimura:2002te}.

\begingroup
\begin{table}[!htb]
\caption{Predicted reach in effective couplings ($g^\prime/g_{SM}$) for
95\% C.L.\ exclusion of right- and left-handed $W^\prime$ bosons at
$\sqrt{S}=$ 7 TeV in 5 fb$^{-1}$ of data.  Positive ($+$ int.) and negative
($-$ int.) interference limits for $W^\prime_L$ are listed separately.
\label{tab:7coupling}}
\begin{ruledtabular}
\begin{tabular}{rrrr}
$W^\prime$ mass & Right & Left ($+$int.) & Left ($-$int.) \\\hline 
 0.50 TeV& 0.16 & 0.20 & 0.21 \\
 0.75 TeV& 0.20 & 0.26 & 0.27  \\  
 1.00 TeV& 0.26 & 0.30 & 0.33  \\  
 1.25 TeV& 0.44 & 0.51 & 0.58  \\  
 1.50 TeV& 0.65 & 0.72 & 0.84  \\  
 1.75 TeV& 0.89 & 0.85 & 1.04  \\  
 2.00 TeV& 1.84 & 1.76 & 2.24  \\  
 2.25 TeV& 3.20 & 3.29 & 3.68  \\  
 2.50 TeV& 5.99 & 5.96 & 6.67  \\  
 2.75 TeV& 11.59 & 10.58 & 11.83  \\ 
 3.00 TeV& 22.15 & 16.43 & 23.12  
% 3.25 TeV& 44.47 & 31.06 & 43.71  \\ 
% 3.50 TeV& 92.41 & 63.24 & 88.93  \\ 
\end{tabular} 
\end{ruledtabular}
\end{table}
\endgroup

\begin{figure}
\includegraphics[width=\columnwidth]{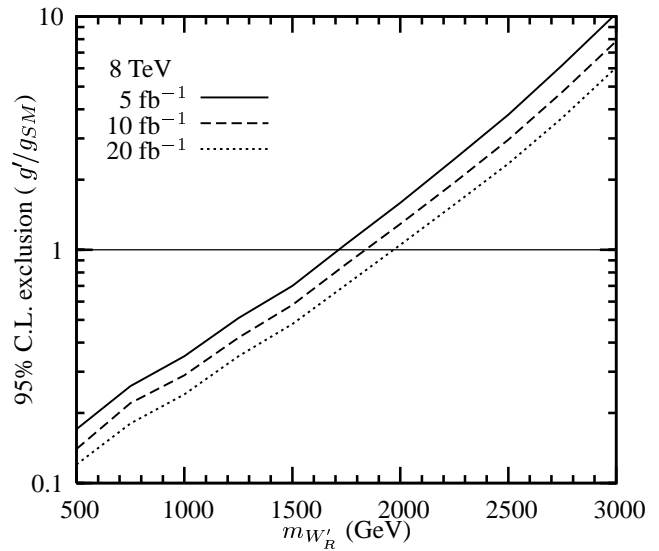}
\caption{Expected 95\% C.L.\ exclusion limit for $g^\prime/g_{SM}$ for a
  right-handed $W^\prime_R$ boson at $\sqrt{S}=8$ TeV, with 5, 10, or 20
 fb$^{-1}$ of data.
\label{fig:RequiredCoupRE}}
\end{figure}

\begin{figure}
\includegraphics[width=\columnwidth]{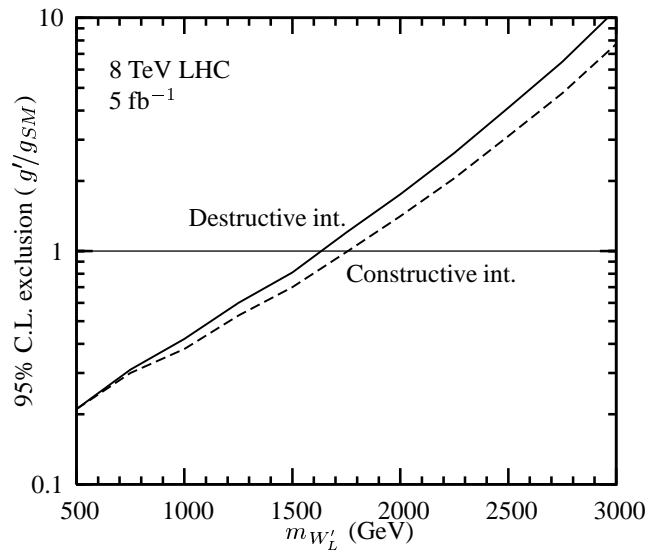}
\caption{Expected 95\% C.L.\ exclusion limit for $g^\prime/g_{SM}$ for a
  left-handed $W^\prime_L$ boson at $\sqrt{S}=8$ TeV and 5 fb$^{-1}$ of data
  that interferes constructively or destructively with standard model
  single-top-quark production.
\label{fig:RequiredCoupLE}}
\end{figure}

In moving to 8~TeV at the LHC we expect a slightly improved mass reach
for SM-like $W^\prime$ bosons due to the additional available energy,
and improved reach in $g^\prime/g_{SM}$ from the increased signal
cross section.  However, we see in Table \ref{tab:8coupling} and
Figs.\ \ref{fig:RequiredCoupRE} and \ref{fig:RequiredCoupLE} that at
5~$\fbi$ integrated luminosity, the reach is actually slightly worse
than at 7~TeV.  In general, the search below 2~TeV becomes more
difficult because the gluon-initiated backgrounds ($Wjj$ and $t\bar
t$) grow faster with collider energy than the quark-initiated signal.
In addition, the acceptance of the right-handed $W^\prime_R$ are
reduced when compared to left-handed $W^\prime_L$ due to the kinematic
distributions discussed in Sec.\ \ref{sec:kinematics}.  The greatest
improvement in searches at the 8 TeV LHC come for masses above 2~TeV,
and with the accumulation of additional data.  In general, the limit on
$g^\prime/g_{SM}$ improves roughly as the fourth-root of the
integrated luminosity ($L^{1/4}$).  Hence, the LHC will improve upon
limits from 7~TeV once the full data sample is accumulated.

\begingroup
\begin{table}[!htb]
\caption{Predicted reach in effective couplings ($g^\prime/g_{SM}$) for
95\% C.L.\ exclusion of right- and left-handed $W^\prime$ bosons at
$\sqrt{S}=$ 8 TeV in 5 fb$^{-1}$ of data.  Positive ($+$ int.) and negative
($-$ int.) interference limits for $W^\prime_L$ are listed separately.
    \label{tab:8coupling}}
\begin{ruledtabular}
\begin{tabular}{rrrr}
$W^\prime$ mass & Right & Left ($+$int.) & Left ($-$int.) \\\hline 
 0.50 TeV& 0.17 & 0.21 & 0.21  \\ 
 0.75 TeV& 0.26 & 0.30 & 0.31  \\ 
 1.00 TeV& 0.35 & 0.38 & 0.42  \\ 
 1.25 TeV& 0.51 & 0.53 & 0.60  \\ 
 1.50 TeV& 0.70 & 0.70 & 0.81  \\ 
 1.75 TeV& 1.06 & 0.99 & 1.20  \\ 
 2.00 TeV& 1.59 & 1.41 & 1.75  \\ 
 2.25 TeV& 2.44 & 2.05 & 2.63  \\ 
 2.50 TeV& 3.79 & 3.11 & 4.11  \\ 
 2.75 TeV& 6.19 & 4.75 & 6.47  \\ 
 3.00 TeV& 10.29 & 7.71 & 10.84 
% 3.25 TeV& 16.85 & 11.64 & 16.80  \\ 
% 3.50 TeV& 28.32 & 18.96 & 28.13  \\ 
\end{tabular} 
\end{ruledtabular}
\end{table}
\endgroup

The reach we predict with the cuts shown here represents an improvement
over the similar ATLAS $W^\prime$ analysis of Ref.~\cite{Aad:2012ej},
which uses fixed cuts for all masses.  This is due to the fact that as the
$W^\prime$ mass increases, the lead jet cut is the most effective cut
in removing background contamination.  Our results are similar to the
CMS analysis in \cite{:2012sc}, although we stress the
coupling-dependence as being critical for comparison with theory.
The top mass cut coupled with our scaling leading jet cut has the
potential to improve on the current CMS results by eliminating a large
portion of the $Wjj$ background and a significant portion of the $t
\bar t$ background.

%%%%%%%%%%%%%%%%%%%%%%%%%%%%%%%%%%%%%%%%%%%%%%%%%%%%%%%%%%%%%%%%%%%%%%%%%%

\section{Conclusions}
\label{sec:concl}

In this paper we investigate the reach in mass and coupling for
arbitrary $W^\prime$ boson models where the $W^\prime$ decays to $tb$.
For purely right-handed $W^\prime_R$ bosons, the 7 TeV run of the LHC
could exclude standard model-like ($g^\prime_R=g_{SM}$) bosons of mass
up to 1800 GeV with 5~$\fbi$ of integrated luminosity.  Larger
backgrounds at 8 TeV lead to a lower reach of 1700~GeV with 5~$\fbi$
of data, but a combined analysis of 20~$\fbi$ may exclude up to 1950
GeV.  For left-handed $W^\prime_L$ bosons, a reach of 1750--1900 GeV
is possible with 5~$\fbi$ depending on the sign of interference with
standard model single-top-quark production.  With a combined run of
20~$\fbi$ at 8 TeV, exclusions of 1900--2050 GeV are possible, with
the highest exclusion when the $W^\prime_L$ interacts constructively
with the standard model $W$ boson.

In addition to $W^\prime$ mass reach for standard model-like couplings
we demonstrate the reach as a function of relative effective coupling
$g^\prime/g_{SM}$.  We find that for masses near 1~TeV, 95\% C.L.\
exclusion limits can be set around $g^\prime/g_{SM}\sim 0.2$--$0.3$.
In addition, near 2.5~TeV, limits can be set on couplings below 3.
These coupling-dependent limits are important, because they cover a
large range of perturbative models.  For example, models in which
there are mixtures of multiple SU(2)$_L$ there are constraints on their
couplings due to the measurement of $g_{SM}$:
\begin{equation}
\frac{1}{g_1^2} + \frac{1}{g_2^2} + \cdots + \frac{1}{g_n^2} =
\frac{1}{g_{SM}^2} \approx \frac{1}{0.427}\,.
\end{equation}
This implies $1.02 g_{SM} < g_{1,2,\ldots} < \sqrt{4\pi}$ in these
perturbative models.  Hence, there will always be at least one
$W^\prime$ boson with $0.187 < g^\prime/g_{SM} < 5.34$
\cite{Sullivan:2003xy}.  While many theoretical models have a
preference for $g^\prime/g_{SM} \sim 1$, the coupling dependent limit
catches most of them.

In some models the $W^\prime$ boson might decay through other
recognizable channels with the same final state, e.g., $W^\prime \to
WZ\to Wb\bar b$ or $W^\prime \to WH\to Wb\bar b$.  These channels
would both be detectable and have similar backgrounds to the
single-top signal, but some of the cuts used in the analysis would no
longer apply.  Instead of a loose cut on the top-quark mass and a
large leading jet $E_T$ requirement, a relatively tight cut could be
placed on the two jets to reconstruct the Higgs \cite{:2012gu,:2012gk}
or $Z$ boson mass.  Even a relaxation of the loose top-mass cut would
allow enough access to these channels to say something about many of
the models listed in Sec.\ \ref{sec:intro}.

If a $W^\prime$ boson is discovered using these methods, the next
logical step would be to establish its chirality.  This can be
accomplished by looking at the angular correlations between the
charged lepton and the initial-state down quark in the reference frame
of the top quark \cite{Sullivan:2002jt}.  At the LHC this is
corresponds to a broadening of the pseudorapidity distribution of the
lepton.  Other kinematic variables shown in Sec.\ \ref{sec:kinematics}
may also be used.

While this paper discusses exclusion, the reach for discovery of a
$W^\prime$ boson scales like $(g^\prime)^2$ --- i.e. the discovery
reach curves are 1.6$\times$ the exclusion curves of Figs.\
\ref{fig:RequiredCoupRS}--\ref{fig:RequiredCoupLE}.  Hence, the model
independent search for $W^\prime$ bosons presented here has the
potential to explore the parameter space of most charged vector
current models that have been proposed within the past few years.

\begin{acknowledgments}
  This work is supported by the U.S.\ Department of Energy under
  Contract No.\ DE-SC0008347.
\end{acknowledgments}

%%%%%%%%%%%%%%%%%%%%%%%%%%%%%%%%%%%%%%%%%%%%%%%%%%%%%%%%%%%%%%%%%%%%%%%%%%

\appendix
\section*{Appendix}
\label{sec:app}

We summarize here the inclusive cross sections plus theoretical
uncertainties for the single-top-quark final state of $W^\prime$
production at the Large Hadron Collider.  Cross sections are
calculated for $t\bar b$ and $\bar t b$ separately, for both 7~TeV and
8~TeV $pp$ colliders.  Cross sections at 7~TeV for right-handed
$W^\prime_R$ bosons are listed by mass at leading order and NLO in
femtobarns in Table \ref{tab:sigwprs}; and left-handed $W^\prime_L$
bosons are listed by mass at leading order and NLO in femtobarns in
Table \ref{tab:sigwpls}.  Note, the left-handed cross sections assume
\textit{no interference} with the standard model production process.
See Sec.\ \ref{sec:sim} for a description of how we use the
left-handed normalization.  Cross sections at 8~TeV are listed for
$W^\prime_R$ and $W^\prime_L$ in Tables \ref{tab:sigwpre} and
\ref{tab:sigwple}, respectively.

\begingroup
\begin{table}
  \caption{LO and NLO cross sections in (fb) for $pp\to W^{\prime}_{R}\to 
    t\bar b\, (\bar t b)$ at the LHC, $\sqrt{S}=7$ TeV, where the decay to
    leptons is not allowed.  NLO includes all theoretical uncertainties
    listed in the text, and is dominated by PDF uncertainties.
    \label{tab:sigwprs}}
\begin{ruledtabular}
\begin{tabular}{rrrr@{}lr@{}l}
Mass (GeV)&$\sigma^t_{\mathrm LO}\;\mathrm{(fb)}$
&$\sigma^{\bar t}_{\mathrm LO}\;\mathrm{(fb)}$
&\multicolumn{2}{c}{$\sigma^t_{\mathrm NLO}\;\mathrm{(fb)}$}
&\multicolumn{2}{c}{$\sigma^{\bar t}_{\mathrm NLO}\;\mathrm{(fb)}$} \\\hline
500 & 29300 & 13200 & 37600 & $^{+1900}_{-2400}$ & 17100 & $^{+1390}_{-940}$ \\
750 & 6340 & 2400 & 7810 & $^{+550}_{-420}$ & 3090 & $^{+270}_{-250}$ \\
1000 & 1800 & 592 & 2160 & $^{+170}_{-200}$ & 760 & $^{+86}_{-81}$ \\
1250 & 589 & 174 & 689 & $^{+53}_{-71}$ & 225 & $^{+29}_{-29}$ \\
1500 & 209 & 57.2 & 237 & $^{+26}_{-27}$ & 75.1 & $^{+11.7}_{-10.7}$ \\
1750 & 78.2 & 20.4 & 86.5 & $^{+10.1}_{-11.5}$ & 27.4 & $^{+5.1}_{-4.3}$ \\
2000 & 30.2 & 7.80 & 32.7 & $^{+4.5}_{-4.8}$ & 10.9 & $^{+2.2}_{-2.0}$ \\
2250 & 12.1 & 3.18 & 12.9 & $^{+2.2}_{-1.9}$ & 4.57 & $^{+1.03}_{-0.90}$ \\
2500 & 5.07 & 1.40 & 5.47 & $^{+1.01}_{-0.94}$ & 2.06 & $^{+0.49}_{-0.42}$ \\
2750 & 2.28 & 0.68 & 2.55 & $^{+0.49}_{-0.39}$ & 1.01 & $^{+0.23}_{-0.19}$ \\
3000 & 1.13 & 0.36 & 1.34 & $^{+0.23}_{-0.18}$ & 0.55 & $^{+0.11}_{-0.09}$ \\
3250 & 0.63 & 0.22 & 0.80 & $^{+0.11}_{-0.09}$ & 0.33 & $^{+0.05}_{-0.05}$ \\
3500 & 0.40 & 0.14 & 0.52 & $^{+0.06}_{-0.05}$ & 0.21 & $^{+0.03}_{-0.02}$
\end{tabular}
\end{ruledtabular}
\end{table}
\endgroup

\begingroup
\begin{table}
  \caption{LO and NLO cross sections in (fb) for $pp\to W^{\prime}_{L}\to 
    t\bar b\, (\bar t b)$ at the LHC, $\sqrt{S}=7$ TeV, where the decay to
    leptons is allowed, but no interference is included (see text).  NLO
    includes all theoretical uncertainties listed in the text, and is
    dominated by PDF uncertainties.
    \label{tab:sigwpls}}
\begin{ruledtabular}
\begin{tabular}{rrrr@{}lr@{}l}
Mass (GeV)&$\sigma^t_{\mathrm LO}\;\mathrm{(fb)}$
&$\sigma^{\bar t}_{\mathrm LO}\;\mathrm{(fb)}$
&\multicolumn{2}{c}{$\sigma^t_{\mathrm NLO}\;\mathrm{(fb)}$}
&\multicolumn{2}{c}{$\sigma^{\bar t}_{\mathrm NLO}\;\mathrm{(fb)}$} \\\hline
500 & 21500 & 9700 & 27800 & $^{+1600}_{-1100}$ & 12800 & $^{+800}_{-800}$ \\
750 & 4720 & 1790 & 5870 & $^{+410}_{-320}$ & 2330 & $^{+200}_{-190}$ \\
1000 & 1350 & 445 & 1630 & $^{+110}_{-140}$ & 575 & $^{+74}_{-54}$ \\
1250 & 444 & 132 & 522 & $^{+46}_{-46}$ & 172 & $^{+23}_{-20}$ \\
1500 & 159 & 43.9 & 182 & $^{+20}_{-19}$ & 58.2 & $^{+8.8}_{-8.3}$ \\
1750 & 60.1 & 15.9 & 67.4 & $^{+7.3}_{-8.6}$ & 21.6 & $^{+4.0}_{-3.3}$ \\
2000 & 23.6 & 6.23 & 26.0 & $^{+3.4}_{-3.8}$ & 8.71 & $^{+1.68}_{-1.59}$ \\
2250 & 9.70 & 2.62 & 10.6 & $^{+1.7}_{-1.6}$ & 3.77 & $^{+0.78}_{-0.74}$ \\
2500 & 4.21 & 1.20 & 4.67 & $^{+0.75}_{-0.75}$ & 1.76 & $^{+0.378}_{-0.33}$ \\
2750 & 1.98 & 0.60 & 2.27 & $^{+0.38}_{-0.32}$ & 0.90 & $^{+0.17}_{-0.17}$ \\
3000 & 1.03 & 0.34 & 1.25 & $^{+0.20}_{-0.14}$ & 0.50 & $^{+0.09}_{-0.07}$ \\
3250 & 0.60 & 0.21 & 0.77 & $^{+0.09}_{-0.08}$ & 0.31 & $^{+0.05}_{-0.04}$ \\
3500 & 0.39 & 0.14 & 0.51 & $^{+0.05}_{-0.04}$ & 0.21 & $^{+0.02}_{-0.02}$
\end{tabular}
\end{ruledtabular}
\end{table}
\endgroup

\begingroup
\begin{table}
  \caption{LO and NLO cross sections in (fb) for $pp\to W^{\prime}_{R}\to 
    t\bar b\, (\bar t b)$ at the LHC, $\sqrt{S}=8$ TeV, where the decay to
    leptons is not allowed.  NLO includes all theoretical uncertainties
    listed in the text, and is dominated by PDF uncertainties.
    \label{tab:sigwpre}}
\begin{ruledtabular}
\begin{tabular}{rrrr@{}lr@{}l}
Mass (GeV)&$\sigma^t_{\mathrm LO}\;\mathrm{(fb)}$
&$\sigma^{\bar t}_{\mathrm LO}\;\mathrm{(fb)}$
&\multicolumn{2}{c}{$\sigma^t_{\mathrm NLO}\;\mathrm{(fb)}$}
&\multicolumn{2}{c}{$\sigma^{\bar t}_{\mathrm NLO}\;\mathrm{(fb)}$} \\\hline
500 & 36800 & 17400 & 47300& $^{+2300}_{-2800}$ & 22500 & $^{+2000}_{-1200}$ \\
750 & 8370 & 3370 & 10400 & $^{+740}_{-500}$ & 4300 & $^{+470}_{-210}$ \\
1000 & 2520 & 888 & 3080 & $^{+190}_{-293}$ & 1130 & $^{+140}_{-97}$ \\
1250 & 888 & 280 & 1050 & $^{+74}_{-97}$ & 359 & $^{+44}_{-36}$ \\
1500 & 342 & 99.0 & 396 & $^{+36}_{-44}$ & 128 & $^{+19}_{-16}$ \\
1750 & 140 & 38.0 & 157 & $^{+17}_{-16}$ & 50.0 & $^{+7.7}_{-7.6}$ \\
2000 & 59.0 & 15.5 & 65.1 & $^{+8.0}_{-8.1}$ & 20.8 & $^{+3.8}_{-3.4}$ \\
2250 & 25.7 & 6.68 & 27.8 & $^{+3.9}_{-4.1}$ & 9.24 & $^{+1.86}_{-1.68}$ \\
2500 & 11.5 & 3.03 & 12.3 & $^{+2.0}_{-1.8}$ & 4.32 & $^{+0.90}_{-0.86}$ \\
2750 & 5.31 & 1.46 & 5.69 & $^{+1.06}_{-0.88}$ & 2.13 & $^{+0.46}_{-0.44}$ \\
3000 & 2.58 & 0.75 & 2.83 & $^{+0.52}_{-0.46}$ & 1.11 & $^{+0.24}_{-0.22}$ \\
3250 & 1.34 & 0.42 & 1.54 & $^{+0.26}_{-0.24}$ & 0.62 & $^{+0.13}_{-0.11}$ \\
3500 & 0.76 & 0.26 & 0.92 & $^{+0.14}_{-0.11}$ & 0.38 & $^{+0.06}_{-0.06}$
\end{tabular}
\end{ruledtabular}
\end{table}
\endgroup

\begingroup
\begin{table}
  \caption{LO and NLO cross sections in (fb) for $pp\to W^{\prime}_{L}\to 
    t\bar b\, (\bar t b)$ at the LHC, $\sqrt{S}=8$ TeV, where the decay to
    leptons is allowed, but no interference is included (see text).  NLO
    includes all theoretical uncertainties listed in the text, and is
    dominated by PDF uncertainties.
    \label{tab:sigwple}}
\begin{ruledtabular}
\begin{tabular}{rrrr@{}lr@{}l}
Mass (GeV)&$\sigma^t_{\mathrm LO}\;\mathrm{(fb)}$
&$\sigma^{\bar t}_{\mathrm LO}\;\mathrm{(fb)}$
&\multicolumn{2}{c}{$\sigma^t_{\mathrm NLO}\;\mathrm{(fb)}$}
&\multicolumn{2}{c}{$\sigma^{\bar t}_{\mathrm NLO}\;\mathrm{(fb)}$} \\\hline
500 & 27000 & 12800 & 35300 & $^{+1300}_{-2300}$ & 16800 & $^{+910}_{-1200}$ \\
750 & 6230 & 2510 & 7780 & $^{+600}_{-380}$ & 3260 & $^{+230}_{-290}$ \\
1000 & 1890 & 667 & 2320 & $^{+170}_{-200}$ & 860 & $^{+96}_{-78}$ \\
1250 & 668 & 212 & 800 & $^{+52}_{-79}$ & 273 & $^{+37}_{-26}$ \\
1500 & 259 & 75.5 & 302 & $^{+29}_{-30}$ & 98.0 & $^{+16.2}_{-10.9}$ \\
1750 & 106 & 29.3 & 122 & $^{+11}_{-14}$ & 38.8 & $^{+6.1}_{-5.6}$ \\
2000 & 45.5 & 12.2 & 50.9 & $^{+5.7}_{-6.5}$ & 16.5 & $^{+2.7}_{-2.8}$ \\
2250 & 20.1 & 5.34 & 22.1 & $^{+3.0}_{-2.9}$ & 7.41 & $^{+1.45}_{-1.26}$ \\
2500 & 9.17 & 2.49 & 10.0 & $^{+1.5}_{-1.5}$ & 3.55 & $^{+0.71}_{-0.66}$ \\
2750 & 4.37 & 1.24 & 4.79 & $^{+0.81}_{-0.67}$ & 1.80 & $^{+0.38}_{-0.33}$ \\
3000 & 2.20 & 0.66 & 2.48 & $^{+0.41}_{-0.35}$ & 0.98 & $^{+0.19}_{-0.18}$ \\
3250 & 1.19 & 0.38 & 1.40 & $^{+0.21}_{-0.20}$ & 0.57 & $^{+0.11}_{-0.09}$ \\
3500 & 0.70 & 0.24 & 0.86 & $^{+0.12}_{-0.09}$ & 0.36 & $^{+0.05}_{-0.05}$
\end{tabular}
\end{ruledtabular}
\end{table}
\endgroup

Uncertainties in the Tables are listed in femtobarns at NLO.  The
cross section uncertainties are completely dominated by the errors in
the CTEQ 6.6 parton distribution functions \cite{Nadolsky:2008zw}, and
are calculated using the standard Modified Tolerance Method
\cite{Sullivan:2002jt}.  In order of importance, other uncertainties
included are estimates of higher-order effects evaluated by scale
variation, and current measurements of the coupling $\alpha_s$, and
the top quark mass \cite{Beringer:1900zz}.

%%%%%%%%%%%%%%%%%%%%%%%%%%%%%%%%%%%%%%%%%%%%%%%%%%%%%%%%%%%%%%%%%%%%%%%%%%

\end{document}